\documentclass[twocolumn,showpacs,preprintnumbers,amsmath,amssymb]{revtex4}

\usepackage{graphicx}
\usepackage{dcolumn}
\usepackage{bm}

\begin{document}

\title{Angle Dependence of Landau Level Spectrum in Twisted Bilayer Graphene}

\author{Min-Young Choi,~~Young-Hwan Hyun, and~~Yoonbai Kim}
\affiliation{Department of Physics, BK21 Physics Research
Division, and Institute of Basic Science\\
Sungkyunkwan University, Suwon 440-746, Korea}
\email{mychoi22, yhhyun, yoonbai@skku.edu}


\begin{abstract}
In the context of the low energy effective theory, the exact Landau level
spectrum of quasiparticles in twisted bilayer graphene with small twist angle
is analytically obtained by spheroidal eigenvalues.
We analyze the dependence of the Landau levels on the twist
angle to find the points,
where the two-fold degeneracy for twist angles is lifted in the nonzero modes
and below/above which massive/massless fermion pictures become valid.
In the perpendicular
magnetic field of 10\,T, the degeneracy is removed at
$\theta_{{\rm deg}}\sim 3^\circ$
for a few low levels, specifically, $\theta_{\rm deg}\simeq 2.56^\circ$ for the first pair of nonzero levels and
$\theta_{\rm deg}\simeq 3.50^\circ$  for the next pair.  Massive quasiparticle appears at
$\theta<\theta_{{\rm c}}\simeq 1.17^\circ$ in 10\,T,
which match perfectly with the recent experimental results.
Since our analysis is applicable to the cases of arbitrary constant magnetic
fields, we make predictions for the same experiment performed in
arbitrary constant magnetic fields, e.g., for $B=40$\,T we get
$\theta_{\rm c}\simeq 2.34^\circ$ and the
sequence of angles $\theta_{\rm deg} = 5.11, 7.01, 8.42,\cdots$ for the pairs of nonzero energy levels.
The symmetry restoration mechanism behind the massive/massless transition is conjectured to be a tunneling (instanton) in momentum space.

\end{abstract}
\maketitle

\section{Introduction}

One of the most prominent hallmarks of the existence of massless Dirac
fermions in monolayer graphene was the experimental discovery of an unusual
quantum Hall effect \cite{QHE1}. In other words, the observation of the Landau
level (LL) pattern of the massless
Dirac fermions, in the presence of a perpendicular magnetic field $B$, $E_{n}\sim\pm\sqrt{Bn}$ ($n=0,1,2,\cdots$), confirmed the characteristic band structure of graphene, long
after its theoretical prediction \cite{Wall}.

The quasiparticles in bilayer graphene (BLG) of Bernal stacking are massive
and described by the following Hamiltonian. For a given valley, say ${\bf K}$,
in the low energy continuum limit, it is
\begin{align}
H^0_{\bf K} = -\frac{\hbar^2}{m}\begin{pmatrix} 0 & \partial^2 \\
\bar\partial^2 & 0 \end{pmatrix}~, \label{untwisted}
\end{align}
where $m$ is the effective mass of the quasiparticle in BLG and we have
introduced complex coordinates $z \equiv(x+iy)/{\sqrt2}$ on the graphene plane. The derivative operators are defined as
$\partial\equiv \partial/\partial z$ and $\bar\partial\equiv \partial/\partial{\bar z}$, respectively.
The LL spectrum of BLG is thus different from that of monolayer graphene
and is given by eigenvalues of the following Hamiltonian in a magnetic field $B$:
\begin{align}
H^{B}_{\bf K} = -\hbar\omega \begin{pmatrix} 0 & \hat a^{\dag 2} \\ \hat a^2 & 0 \end{pmatrix}~,
\end{align}
that is, by $E_n = \pm\hbar\omega \sqrt{n(n-1)}$ \cite{McF},
where $\omega=eB/m$ is the cyclotron frequency. The lowering/raising operators $\hat a$/$\hat a^\dag$ satisfy the usual
harmonic oscillator algebra and are given by suitable combinations of covariant derivatives representing the magnetic field $B$.

When graphene layers stack together, the interlayer couplings significantly
change the nature of quasiparticles, like the BLG case considered above.
Surprisingly, massless Dirac fermions survive the stacking in multilayer
structure grown on SiC \cite{massless}. The main reason for this stunning
effect is thought to be the decoupling of twisted layers
\cite{dosSantos},\cite{twistagain}-\cite{twistagain8}. Persistence and properties of massless Dirac fermions at small
twist angles are, however, under some controversy. According to {\it ab initio}
calculations \cite{abinitio}, the decoupling occurs at any values of twist
angle and massless Dirac fermions are essentially those of monolayer graphene. On the other hand, the tight-binding analysis \cite{dosSantos} indicates a strong role played by the interlayer coupling, which considerably affects the nature of quasiparticles. These results are based on the band structure calculation in the absence of magnetic field. In a recent experiment \cite{Luican}, the issue of angle dependence of the LL's in the presence of magnetic field is addressed  by combining scanning tunneling microscopy and LL scanning tunneling spectroscopy. They measured some critical angles at which two-fold degeneracy due to the presence of two massless Dirac fermions is lifted  and where the massless Dirac fermion picture breaks down. Specifically, the degeneracy can be seen at angles above roughly
$\theta_{{\rm deg}}\sim 3^\circ$ for a few low levels in the magnetic field of 10\,T, and the critical value of twist angle below/above which massive/massless LL spectrum is shown is $\theta_{\rm c}\simeq 1.16^\circ$ for 10\,T. An effective Hamiltonian to obtain the corresponding LL's is suggested recently, but only the zero-modes are analytically investigated \cite{deGail}. In order to study the angle dependence we should have controls over the nonzero-modes.

In this paper, we exactly solve the LL spectrum of the Hamiltonian
proposed in \cite{deGail} to give the concrete values of angles,
which show very precise agreements with the measured values for the magnetic
field 10\,T \cite{Luican}.
Since the LL spectrum is obtained in analytic form, we can predict
$\theta_{\rm c}$ and $\theta_{{\rm deg}}$ for every nonzero pair of the LL's in the presence of arbitrary constant magnetic field (under a plausible assumption on the physical continuity of the LL spectrum as functions of twist angle about which we discuss in Appendix\,\ref{appA}). Furthermore, our analytic result for LL's \eqref{Eei} smoothly interpolates between the spectra known before for
massive/massless quasiparticles in BLG, and allows one to get systematic power series corrections for both of two sides of the spectrum.

A natural question one can ask here is: ``What is the symmetry restoration mechanism behind the transition between the massive/massless spectra?'' Based on our exact results, we anticipate that the non-perturbative symmetry restoration mechanism is a tunneling (instanton) in the momentum (reciprocal) space \cite{Jaffe}.

The remaining parts of this paper are organized as follows. In Sec.~\ref{sec2}, we briefly review the construction of the low-energy effective Hamiltonian for charged quasiparticles in twisted BLG \cite{deGail} and solve the associated LL problem analytically by invoking a differential-equation representation of the eigenproblem. In Sec.~\ref{sec3}, we reveal the distinction between two asymptotic regions in LL spectrum, obtained in Sec.~\ref{sec2}, considered as functions of twist angle. Finally,  in Sec.~\ref{sec4}, we wrap this paper up with a short summary and some discussions. An appendix is devoted to discuss the change in the LL spectrum driven by the Fermi speed renormalization as twist angle increases
and to advocate the validity of the analysis made in Sec.~\ref{sec3}.

\section{Exact Landau levels in twisted bilayer graphene}\label{sec2}

\begin{figure}[!ht]
\includegraphics[scale=0.45]{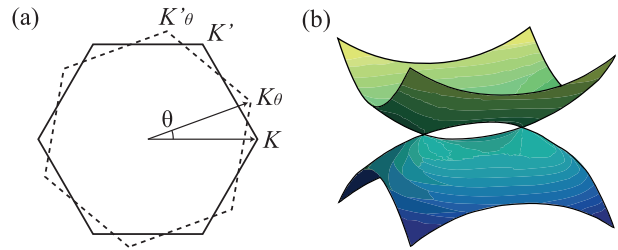}
\caption{\small (Color online) (a) First Brillouin zone for twisted BLG. The first Brillouin zone of the upper layer (dashed hexagon) is rotated by
an angle $\theta$ with respect to that of the lower layer (full hexagon). (b) Low energy band structure near the ${\bf K}$-valley of twisted BLG. The dispersion relation for
quasiparticles is given by $E(k,\bar k)=\pm\frac{\hbar^2}{m(\theta)}|k-\frac{\Delta K}{2}||k+\frac{\Delta K}{2}|$, where $k\equiv\frac{1}{\sqrt2} (k_x+i k_y)$,
$\Delta K \equiv\frac{1}{\sqrt2} (\Delta K_x +i\Delta K_y)$, and $m(\theta)$ is $\theta$-dependent effective mass. If we set $\kappa\equiv k-\frac{\Delta K}{2}$, then we get the massless behavior $E\approx \pm{\sqrt2} \hbar \tilde v_{\rm F}|\kappa|$ near $\kappa=0$,  where $\tilde v_{\rm F}\equiv\frac{\hbar|\Delta K|}{{\sqrt2}\,m(\theta)}$ is the renormalized Fermi speed.}
\label{fig:band}
\end{figure}

Twisted bilayer graphene is a structure specified by a rotational mismatch given by an angle $\theta$
with respect to the perfect Bernal (AB) stacked bilayer graphene. This twisted pattern is not difficult to find but
can be seen on the surface of graphite, for an example. In low energy limit, it shows a drastically different electronic band structure from that of the Bernal-stacked BLG. Its low energy quasiparticles are two massless Dirac fermions rather than one massive fermion, per each valley ({\bf K}/$-${\bf K}) \cite{dosSantos}. For a small $\theta$, the apices of the associated Dirac cones are separated by $|\Delta{\bf K}|=|{\bf K}-{\bf K}_\theta| \simeq 4\pi \theta/3{\sqrt3}\,a_{\rm cc}$ in reciprocal space, where $a_{\rm cc} \simeq 1.42$\,\AA\, is the distance between two adjacent carbon atoms in the hexagonal lattice. A commensurate rotation with a periodic Moir\'e pattern occurs at the angles $\theta_i$:
\begin{align}
\cos\theta_i = \frac{3i^2+3i+\frac12}{3i^2+3i+1}~,\quad i=0,1,2,\cdots,
\end{align}
and the superlattice structure is specified by basis vectors ${\bf t}_1=i{\bf a}_1+(i+1){\bf a}_2$ and ${\bf t}_2 = -(i+1){\bf a}_1+(2i+1){\bf a}_2$, where
${\bf a}_1$, ${\bf a}_2$ are the Bravais lattice basis vectors in the single layer hexagonal lattice \cite{dosSantos}. The lattice constant of the superlattice is given by $|{\bf t}_1|=|{\bf t}_2|=a_{\rm cc}\,{\sqrt{9i^2+9i+3}}$\,. The reciprocal lattice is spanned by
\begin{align}
&{\bf G}_1=\frac{4\pi}{9i^2+9i+3}[(3i+1){\bf a}_1+{\bf a}_2]~,\cr
&{\bf G}_2=\frac{4\pi}{9i^2+9i+3}[-(3i+2){\bf a}_1+(3i+1){\bf a}_2]~.
\end{align}
The first Brillouin zone for twisted BLG is depicted in Fig.~\ref{fig:band}(a) and the corresponding low energy band structure in the ${\bf K}$-valley is shown in Fig.~\ref{fig:band}(b). Electronic properties of twisted BLG and related systems are currently under intensive debates \cite{twistagain}-\cite{twistagain8}.

In a recent paper \cite{deGail}, an effective Hamiltonian for low energy dynamics of twisted bilayer graphene is proposed. Neglecting commensuration effects between two layers, the Hamiltonian describing twisted BLG around the
pair $({\bf K},{\bf K}_\theta)$ reads
\begin{align}
H_{\rm tw}({\bf k})= \begin{pmatrix} H_{\rm D}({\bf k}+\frac{\Delta{\bf K}}{2}) & \sum H_\perp \\ \sum H^\dag_\perp &
H_{\rm D}({\bf k}-\frac{\Delta{\bf K}}{2})\end{pmatrix}~,\label{fullH}
\end{align}
where $H_{\rm D}$ is the Dirac Hamiltonian for monolayer graphene and $H_\perp$ is the hopping matrix between the
two layers. According to the analysis of the Moir\'e pattern in twisted BLG \cite{dosSantos}, for small twist angle $\theta$, there are three different types of $H_\perp$:
\begin{align}
H_\perp\in \left\{ -\tilde t_\perp\begin{pmatrix} 1 & 1 \\ 1 & 1 \end{pmatrix}, -\tilde t_\perp \begin{pmatrix}
 e^{\mp i\frac{2\pi}{3}} & 1 \\ e^{\pm i\frac{2\pi}{3}} & e^{\mp i\frac{2\pi}{3}} \end{pmatrix}  \right\}~,\label{hperp}
\end{align}
where $\tilde t_\perp$ is a coupling parameter which generally depends on $\theta$. The $\theta$-dependence of $\tilde t_\perp$ is, however, very mild and the Slater-Koster calculation performed in Ref.~\cite{dosSantos} indicates that $\tilde t_\perp \simeq 0.4\,\gamma_1$ is nearly a constant, where $\gamma_1\simeq 0.3\,$eV is the nearest interlayer coupling in the untwisted Bernal-stacked BLG. The first one in \eqref{hperp} corresponds to the Fourier component with the crystal momentum ${\bf G}={\bf 0}$ and the second and the third one with ${\bf G}=-{\bf G}_1,-({\bf G}_1+{\bf G}_2)$. The summation in the interlayer coupling term in \eqref{fullH} runs over these three types of the coupling matrices and the other components are suppressed. Due to this interlayer
coupling, there is a slight difference of $1$ meV order between the energies associated with each Dirac points but this second-order effect in perturbation is negligible. Assuming a simplified interlayer coupling under the condition
$\tilde t_\perp \gg \hbar v_{\rm F} |\Delta{\bf K}|$ ($v_{\rm F}$ is the Fermi speed in the monolayer graphene) \footnote{This subtle assumption may be justified by the fact that the Hamiltonian remains in the same topological universality class \cite{deGail}.},
\begin{align}
\sum H_\perp \rightarrow -3\times\frac{5\,{\tilde t_\perp}}{2}\begin{pmatrix} 0 & 1 \\ 0 & 0  \end{pmatrix}~,\label{simplet}
\end{align}
one can obtain an effective two-band Hamiltonian which resembles \eqref{untwisted}
\begin{align}
H^{\rm eff}_{\rm tw}=-\frac{\hbar^2}{m(\theta)}\begin{pmatrix} 0 & \partial^2 -(\overline{\Delta K}/2)^2 \\  \bar\partial^2 - ({\Delta K}/2)^2 & 0  \end{pmatrix}~,\label{twobandH}
\end{align}
where $m(\theta)=15\,\tilde t_\perp/4\,v_{\rm F}^2$ is $\theta$-dependent effective mass due to the $\theta$-dependent $\tilde t_\perp$. In \eqref{simplet} the multiplication by 3 mimics the summation over three possible crystal momentum ${\bf G}$, and another factor $5/2$ is introduced to match the spectrum at $\theta=0$ (Bernal-stacked BLG). For $\theta=0$, the period of the superlattice is infinite and the summation over ${\bf G}$ is overcounting since ${\bf G}_1={\bf G}_2={\bf 0}$. Thus the multiplication by 3 in \eqref{simplet} should be disregarded in this case and the interlayer coupling becomes that of the Bernal-stacked BLG:
\begin{align}
H_\perp(\theta=0)=-\frac{5\,\tilde t_\perp}{2}\begin{pmatrix} 0 & 1 \\ 0 & 0 \end{pmatrix} \simeq \begin{pmatrix} 0 & -\gamma_1 \\ 0 & 0 \end{pmatrix}~.
\end{align}
Then the effective Hamiltonian \eqref{twobandH} goes to \eqref{untwisted}.
Since $|\Delta{\bf K}|$ is proportional to twist angle $\theta$, the two-band approximation made here is restricted to be applicable for very small angles only. Nevertheless, we believe that the LL spectrum given by \eqref{Eei} below is smoothly connected to the spectrum for larger angles, and the analysis made in Sec.~\ref{sec3} is trustable. The multiplicative factor in \eqref{simplet} also plays a crucial role in this context.  We relegate the discussion of this issue to the appendix of this paper.

In the presence of a perpendicular magnetic field, $B$, the Hamiltonian is written in the form
\begin{align}
H_{B} = -\hbar\omega(\theta) \begin{pmatrix} 0 & \hat a^{\dag 2} -\bar\beta^2 \\
\hat a^2 - \beta^2\end{pmatrix}~,\label{Htwist}
\end{align}
where $\omega(\theta)\equiv eB/m(\theta)$. The lowering/raising operators $\hat a$/$\hat a^\dag$ are given in terms of the covariant derivatives $D\equiv
\partial +\frac{ie}{\hbar}A(z,\bar z)$ / $\bar D\equiv \bar\partial +\frac{ie}{\hbar}\bar A(z,\bar z)$, and $A\equiv\frac{1}{\sqrt2}(A_x-iA_y)$ is a complex vector potential. With the gauge choice $A=-\frac{i}2 B\bar z$, we specifically have
\begin{align}
\hat a= \sqrt{\frac{\hbar}{eB}} \bar D,\quad \hat a^\dag = -\sqrt{\frac{\hbar}{eB}} D~,
\end{align}
which satisfy $[\hat a,\hat a]=[\hat a^\dag,\hat a^\dag]=0$ and $[\hat a,\hat a^\dag]=1$. $\beta$ is a complex parameter proportional to $\Delta K$ as
\begin{align}
\beta \equiv \sqrt{\frac{\hbar}{4eB}}\, \Delta K = \sqrt{\frac{\hbar}{8eB}}\,(\Delta K_x +i\Delta K_y)~.
\end{align}
We can make $\beta$ real-valued simply by rotating the coordinates,
\begin{align}
\beta = \sqrt{\frac{\hbar}{8eB}}\,|\Delta{\bf K}|~.\label{defbeta}
\end{align}
After the rotation, for $\psi=(\psi_1\,\psi_2)^T$, the eigenvalue problem
for the Hamiltonian \eqref{Htwist},
%
$H_{B} \psi = E\,\psi ,$
%
reduces to the following one-component problem:
\begin{align}
\left(\hat a^{\dag 2}-\beta^2\right)(\hat a^2 - \beta^2)\psi_1 = \lambda\psi_1~,\label{reduced}
\end{align}
where $\lambda = [E/\hbar\omega(\theta)]^2$. The remaining component $\psi_2$ can be expressed in terms of $\psi_1$, $\lambda$, and the lowering operator $\hat a$ (except the case $\lambda=0$). Using the anti-holomorphic representation \cite{Faddeev},
\begin{align}
\hat a^\dag \mapsto x~,\quad \hat a \mapsto \frac{d}{dx}~,
\end{align}
the eigenvalue problem \eqref{reduced} is expressed as a second order ordinary differential equation,
\begin{align}
(x^2-\beta^2)(u''-\beta^2 u)-\lambda u = 0~.
\end{align}
Here, the variable $x$ is not the coordinate in the graphene plane and therefore the function $u(x)$ representing
eigenstates are not the wave function in coordinate space.
By rescaling $x\mapsto \beta x$ and setting $u(x)=\sqrt{x^2-1}\, v(x)$, we get the spheroidal equation of
$p=\beta^{2}$ and $q=1$
\begin{align}
[(x^{2}-1)v']'+\left[-p^{2}(x^{2}-1)-\lambda-\frac{q^{2}}{x^{2}-1}\right]v=0
\end{align}
which is a particular case $b=s=0$ of
the confluent Heun's equation: $[(x^2-1)v']'
+[-p^2(x^2-1) +2 p b x-\lambda- \frac{q^2+s^2+2qsx}{x^2-1}]v=0$~\cite{Ronveaux}.

\begin{figure}[!ht]
\includegraphics[scale=0.90]{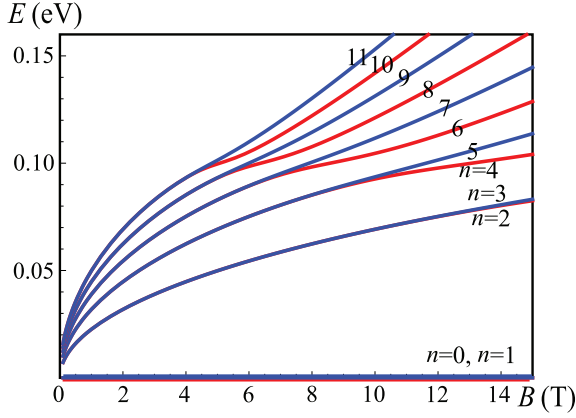}
\caption{\small (Color online) Landau levels of twisted bilayer graphene, given by the spheroidal eigenvalues. The
characteristic energy scale of the van-Hove singularity has been chosen as $\hbar\omega\beta^2=\hbar^2
|\Delta{\bf K}|^2/8m=0.1$\,eV. This scale corresponds to the twist angle of $\theta\simeq 3.27^\circ$.
}
\label{fig:twisted}
\end{figure}

The eigenvalues of the confluent Heun's equations are
$\lambda^{(a)}_{q,s,n-1}(p,b)$, and then
the corresponding LL's are given by the spheroidal
eigenvalues
\begin{align}
  E_n^2=(\hbar \omega(\theta))^2
  \lambda^{(a)}_{1,0,n-1}(\beta^2,0)~,
\label{Eei}
\end{align}
where the superscript $(a)$ stands for `angular'. The LL spectrum for a fixed value of $\beta$ (twist angle) is depicted in Fig.~\ref{fig:twisted}. This analytic result reproduces the numerical calculation performed in Ref.\,\cite{deGail}, except the scale of $B$-dependence due to the multiplicative factor in \eqref{simplet}.

\section{Analysis of spectrum in two asymptotic regions}\label{sec3}

\begin{figure}[!ht]
\includegraphics[scale=1.05]{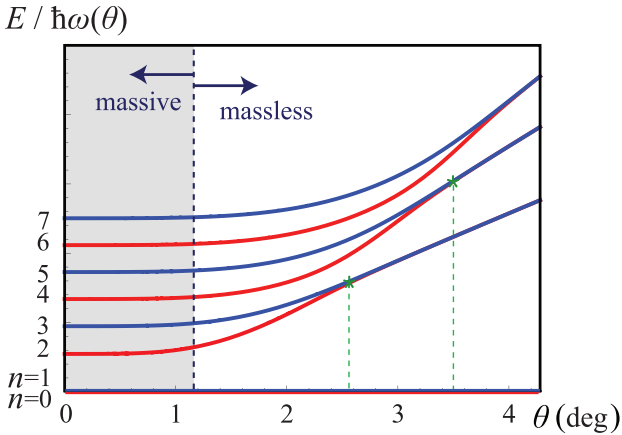}
\caption{\small (Color online) Landau level energies divided by $\theta$-dependent energy scale $\hbar\omega(\theta)$, for a fixed perpendicular magnetic field $B=10$\,T, plotted against twist angle. If the $\theta$-dependence of $\omega(\theta)$ is mild enough, the qualitative behavior of the LL's is the same as $E/\hbar\omega(\theta)$.}
\label{fig:smallth}
\end{figure}

Let us consider the region of the parameter $\beta$ in which the band structure is well-described by two massless Dirac fermions whose
Fermi speed is renormalized to be smaller than that of the monolayer graphene. In this region, the LL's must show a two-fold degeneracy that reflects
the existence of two copies of fermions. Fig.~\ref{fig:twisted} reveals its tail, and Fig.~\ref{fig:smallth} clearly indicates that degeneracy. The lifting of this degeneracy signals breakdown of the description of
electronic bands by two copies of massless fermions and it is caused by the contribution from a saddle point in the band located between the two Dirac cones. Twist-induced van-Hove singularity eventually dominates the spectrum.

Small $\beta$ expansion \cite{Komarov} for the eigenvalues \eqref{Eei} gives,
\begin{align}
  \frac{E_n^{2}}{(\hbar \omega(\theta) )^2} & \approx
     n(n-1)
     -\frac{2n(n-1)}{(2n+1)(2n-3)}\beta^4\cr & ~~
     +\frac{2n(n-1)[n(n-1)(4n(n-1)-39)+63]}{(2n+3)(2n+1)^3(2n-3)^3(2n-5)}
     \beta^8\cr &~~
     +{\cal O}(\beta^{12})~.
\label{sbe}
\end{align}
Untwisted ($\beta=0$) LL spectrum is given by the leading term in \eqref{sbe}. The common factor $n(n-1)$ appears in all the higher order terms in \eqref{sbe} and the modes labeled by $n=0$ and $n=1$ remain to be zero-energy modes even if the corrections in higher powers of $\beta$ are considered. Therefore they are protected. Notice that we are neglecting any other potential lifts of degeneracy due to Zeeman effect, interactions, etc. This protection is due to the topological structure of the band in twisted BLG \cite{deGail}. In twisted BLG, the two Dirac cones are not related by time-reversal symmetry and they are described by the same Berry phase, so that they
are of different topological structure from the Dirac cones in monolayer graphene.

For sufficiently large $\beta$, the asymptotic expansion of the eigenvalues \cite{Komarov} gives
\begin{multline}\label{largeB}
  \frac{E_n^{2}}{(\hbar \omega(\theta) )^2}\approx
  4\left[\frac{n}{2}\right] \beta^2-2\left[\frac{n}{2}\right]^2
  -\left[\frac{n}{2}\right]^3\frac{1}{\beta^2}+\cdots ~,
\end{multline}
where the square bracket denotes the integer part of $n/2$.
The presence of this integer-valued function implies that there are asymptotic degeneracies
between the each even level and the next odd one. The difference between the $2\ell$-th and $(2\ell+1)$-th LL's decreases
exponentially and is vanishing at infinite $\beta$:
\begin{equation}
  \frac{E_{2\ell+1}^2-E_{2\ell}^2}{(\hbar \omega(\theta) )^2}\approx
  \frac{2^{4\ell+3}\beta^{4\ell+2} }{(\ell-1)!\,\ell!} {e}^{-2{\beta^2}}+\cdots \; .\label{asympdeg}
\end{equation}
The exponential factor is independent of $\ell$ but the power of the monomial
function is proportional to $\ell$ so that, in physics experiments, the degeneracy between the adjacent
two levels of lower $\ell$ (or $n$) is lifted for smaller $\beta$ (smaller $\theta$),
which is consistent with Fig.~\ref{fig:smallth}. From now on, we will set the external magnetic field $B=10$\,T, perpendicular to the plane of BLG. Since the value of $\beta$ is large enough up to a very small value of $\theta$ ($\beta\sim1$ for $\theta\sim 1.17^\circ$), we can use the asymptotic formula \eqref{asympdeg} in order to estimate the point on which the two-fold degeneracy is lifted, under the assumption that the $\theta$-dependence of $\omega(\theta)$ (or that of $\tilde t_\perp$) is mild enough.  The values $\beta_{\rm deg}$ for $\ell=1$ where the $n=2$ and $n=3$ levels become non-degenerate and for $\ell=2$ ($n=4,5$) are, from \eqref{asympdeg},
\begin{align}
\frac{2^{4\ell+3}\beta_{\rm deg}^{4\ell+2} }{(\ell-1)!\,\ell!} \mathrm{e}^{-2{\beta_{\rm deg}^2}} \simeq1~,\label{betadg}
\end{align}
and thus
\begin{align}
\beta_{\rm deg}^{(\ell=1)}\simeq 2.18~\quad{\rm and}\quad \beta_{\rm deg}^{(\ell=2)}\simeq 2.99~,
\end{align}
respectively. They correspond to the twisted angles $\theta_{\rm deg}\simeq 2.56^\circ$ and $\simeq 3.50^\circ$ (see the Fig.~\ref{fig:smallth}), in agreement with the measured value, about $3^\circ$, from Ref.\,\cite{Luican}.

As the twist angle $\theta$ decreases, $\beta$ crosses the transition point $\beta_{\rm c}=1$ between the region where the small $\beta$ expansion \eqref{sbe} can be trusted and the region where the large $\beta$ expansion \eqref{largeB} is trustworthy. We will call the region $\beta>\beta_{\rm c}$ massless region and $\beta<\beta_{\rm c}$ massive region, respectively, because of an obvious reason from the behaviors of the LL's in each domain, \eqref{sbe} and \eqref{largeB}. The critical point $\beta_{\rm c}=1$ can roughly  be considered as the point at which the van-Hove singularity eventually starts dominating the spectrum and the description of the band by two massless Dirac fermions breaks down. This critical value $\beta_{\rm c}=1$ corresponds to the twist angle $\theta_{\rm c}\simeq 1.17^\circ$ as mentioned already. The measured value $\theta_{\rm c}^{({\rm measured})} \simeq 1.16^\circ$ \cite{Luican} is very close to our theoretical value, even though there is no exact criterion of fixing the value due to the continuity
of the spectrum.

\begin{figure}[!ht]
\includegraphics[scale=0.75]{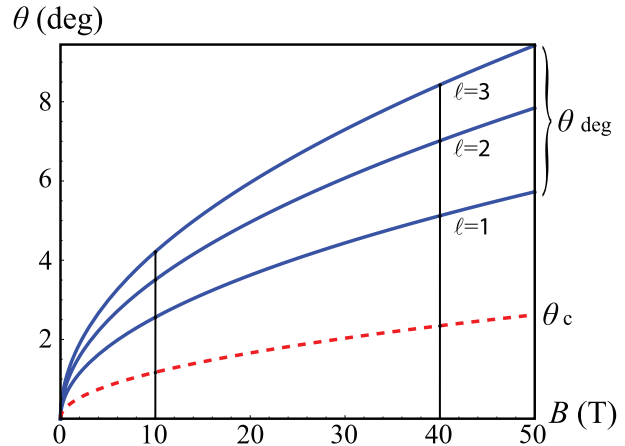}
\caption{\small (Color online) The dependence on $B$ of the various specific angles.}
\label{fig:sqrtB}
\end{figure}

Currently obtainable maximum value of a static magnetic field is about 40\,T.
The critical value $\beta_{\rm c}=1$ and the value of $\beta_{\rm deg}$ for
each $\ell$ in \eqref{betadg} are independent of $B$ and thus the only effect
on the values of various angles for the 40\,T magnetic field is a
multiplication by the factor $2=\sqrt{40/10}$ from \eqref{defbeta} to the
angles for 10\,T. Therefore $\theta_{\rm c}\simeq 2.34^\circ$ and the angles
above which two-fold degeneracy can be seen are as tabulated below, in a high
magnetic field $B=40$\,T.
\begin{center}{
\begin{tabular}{|c|c|c|c|c|}
\hline
$\ell$ & 1 & 2 & 3 &  $\cdots$
\\ \hline
$\theta_{\rm deg}(^\circ)$ & 5.11 & 7.01 & 8.42 & $\cdots$
\\ \hline
\end{tabular}
}\end{center}
This simple dependence of the specific angles $\theta_{\rm deg}$ and $\theta_{\rm c}$ on the magnetic field $B$, that is,
\begin{align}
\theta_{\rm deg,c}(B) = \sqrt{\frac{B}{B_0}}\, \theta_{\rm deg,c}(B_0)~,
\end{align}
where $B_0$ means a reference value of magnetic field, say, $B_0=10$\,T allows us to draw the
Fig.~\ref{fig:sqrtB} to read the angles off
for various values of $B$.

\section{Summary and Discussions}\label{sec4}

To summarize, we solved the eigenvalue problem for LL's in twisted BLG with small twist angles and analyzed the angle dependence of the spectrum in a fixed magnetic field. The corresponding eigenvalues \eqref{Eei} are given by the so-called {\it spheroidal eigenvalues} modulated by an unknown function $\hbar\omega(\theta)$ of twist angle $\theta$, but the modulation does not change the spectrum seriously.  As the results, we got the angle $\theta_{\rm c}$ below/above which the spectrum behaves as LL's for massive/massless fermions and also the angles $\theta_{\rm deg}$ below which the two-fold degeneracies due to the Dirac-point splitting are lifted. The specific values for $\{\theta_{\rm c},\theta_{\rm deg}\}$ for $B<50$\,T can be read off from the Fig.~\ref{fig:sqrtB}. They can be measured in principle by a judicious experiment which combines the scanning tunneling microscopy and the LL scanning tunneling spectroscopy.

The differences between each adjacent energy levels \eqref{asympdeg} show non-perturbative behavior in the expansion parameter,  $1/\beta^2$.
Indeed, the exponential factor in \eqref{asympdeg} gives us a hint about the symmetry restoration mechanism behind the transition described above. It is a typical signal of tunneling mechanism (instanton effect), in this case not in coordinate space but in momentum space \cite{Jaffe}. The situation is analogous to the well-known double-well potential problem
in quantum mechanics. The two Dirac points correspond to classically degenerate ground states, and the van-Hove energy scale plays the role of the height of potential barrier between them. By the tunneling, the Dirac electrons are delocalized in momentum space, thus become localized in coordinate space.

The deformation of the band structure of BLG triggered by twisting is the Dirac-point splitting, as depicted in Fig.~\ref{fig:band}(b). The Dirac-point splitting also happens in untwisted Bernal-stacked BLG by the effect of external magnetic field which is parallel to the graphene plane \cite{Pershoguba,ourtilted}. Therefore, in a tilted magnetic field with respect to the graphene plane, the LL spectrum of untwisted BLG is expected to be the same as \eqref{Eei}, if the parameter $\beta\propto B_\parallel/{\sqrt{B_\perp}}$ represents the effect of the parallel component $B_\parallel$ to the spectrum \cite{ourtilted}. In principle,
the magnetic field component parallel to the plane, which is easy to get by tilting the graphene sample in an external magnetic field, can split the Dirac point and make the LL spectrum doubly-degenerate. It is intriguing to investigate the combined effect of twisting and inclination in a magnetic field, since in reality graphene samples always deviate from the perfect plane by various physical reasons.

\begin{acknowledgments}
The authors benefited from useful discussions with C. Sochichiu.
This work was supported by the Korea Research Foundation Grant funded by
the Korean Government with grant number KRF-2008-313-C00170 and 2011-0011660
 (Y.K.).
\end{acknowledgments}

\begin{appendix}
\section{Renormalized Fermi speed and angle dependence}\label{appA}

The authors of Ref.\,\cite{deGail} managed to get the two-band effective Hamiltonian \eqref{twobandH} by assuming a
simplified form of the interlayer coupling \eqref{simplet} under the condition that $\xi\equiv\hbar v_{\rm F} |\Delta{\bf K}|/\tilde t_\perp \ll 1$, as we reviewed in
Sec.~\ref{sec2}. The reduction from the full Hamiltonian \eqref{fullH} to the effective one \eqref{twobandH} is reminiscent of that of the Bernal-sracked BLG case.
In fact, in addition to the simplified form assumed by the authors of Ref.\,\cite{deGail}, we need a numerical multiplication factor in \eqref{simplet} in order to  connect smoothly the spectrum and the Fermi speed to those for larger angles ($\xi\gg1$) obtained in Ref.~\cite{dosSantos} as we shall see below.

\begin{figure}[!ht]
\includegraphics[scale=0.45]{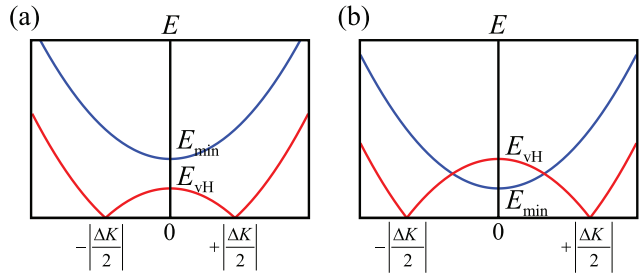}
\caption{\small (Color online) Schematic plots for the (upper half) band structure of the full Hamiltonian \eqref{fullH} along the $\Delta{\bf K}$-direction. (a) The case in which the minimum energy of the upper band is larger than the van-Hove energy. The two-band approximation can be applied without any difficulty. (b) The opposite case. The condition to derive the effective two-band Hamiltonian is not satisfied any more.}
\label{fig:validity}
\end{figure}

The condition of small $\xi$, $\xi\ll1$, used to get the two-band effective Hamiltonian \eqref{twobandH}, can be understood and refined  as follows. The minimum energy of the upper band of the full Hamiltonian, approximately $E_{\rm min} \simeq 15\,\tilde t_\perp/2$, should be much larger than the van-Hove energy of the lower band $E_{\rm vH}=\hbar^2 |\Delta{\bf K}|^2 / 8m(\theta) = \hbar^2 v_{\rm F}^2 |\Delta{\bf K}|^2/ 30\,\tilde t_\perp$, so that they cannot feel the existence of each other (Fig.~\ref{fig:validity} (a)). This refined condition that $\tilde t_\perp \gg \frac1{15} \hbar v_{\rm F} |\Delta{\bf K}|$ ($\xi\ll15$) yields the restriction on the range of angles
\begin{align}
\theta \ll \tilde t_\perp \frac{45{\sqrt3}\,a_{\rm cc}}{4\pi\hbar v_{\rm F}} \simeq 9.80^\circ ~,
\end{align}
%
but this does not tell us sharply how small the twist angle $\theta$ should be. For example, a blind application of our formula to $\theta=8^\circ$ gives a divergent result for a small but finite $B$, say, $B=0.1\,$T. Then, how can we trust the analysis presented in Sec.~\ref{sec3}? Some of the predicted values of $\theta_{\rm c,deg}$ might be in the region where the spectrum is not quite close to that given by \eqref{Eei}.

\begin{figure}[!ht]
\includegraphics[scale=0.9]{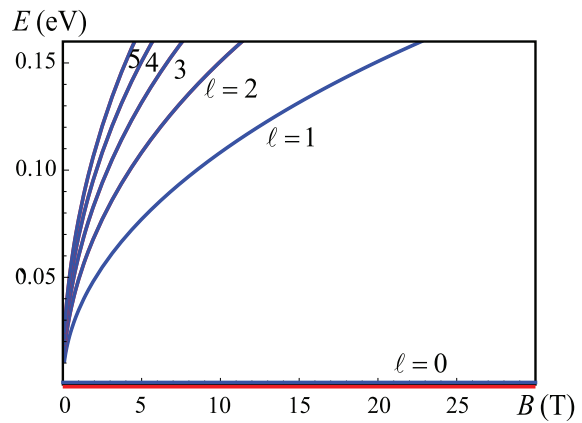}
\caption{\small (Color online) The LL spectrum \eqref{Eei} for $\theta=5^\circ$. Each adjacent levels are degenerate and in the ``massless'' region if the applied magnetic field is not
very high.}
\label{fig:massless}
\end{figure}

Let us recall that for sufficiently large angles, the spectrum given by \eqref{Eei} shows the characteristic LL behavior of massless particles.
The combination of twist angle $\theta\sim5^\circ$ and magnetic field $B=10$\,T for instance corresponds to $\beta\sim 4.26$, in the ``massless region''. Moreover, $\theta\sim5^\circ$ and $\beta_{\rm c}=1$ require $B\sim 182$\,T, a very large field strength. This simple observation means that the LL spectrum of twisted BLG \eqref{Eei} with $\theta \sim 5^\circ$ cannot be pushed to ``massive region'', unless a very strong magnetic field is applied. Let us explain this point in more detail. As the twist angle increases the van-Hove energy, playing the role of a barrier between the two Dirac points, also increases as can be seen in Fig.~\ref{fig:validity} (b). Therefore the ``tunneling'' between the two Dirac points is suppressed and the ${\mathbb Z}_2$ symmetry is broken. Unless a very strong magnetic field is applied, the broken symmetry cannot be restored and the spectrum is described by two massless fermions degenerate in energy, for $\theta \sim 5^\circ$. This fact is clearly shown in Fig.~\ref{fig:massless}, plotted for $\theta=5^\circ$.

Indeed, the masslessness of quasiparticles near one of the two Dirac points was shown in Ref.~\cite{dosSantos} by applying perturbation theory under the opposite condition, that $\xi\gg1$. Their result indicates the Fermi speed renormalization
\begin{align}
 \tilde v_{\rm F}/v_{\rm F}=1-9/\xi^2~,\label{vf1}
\end{align}
where $\tilde v_{\rm F}$ is the renormalized Fermi speed. This result also seems strange since it tells us that the renormalized Fermi speed vanishes at $\xi=3$, corresponding to $\theta\simeq 2^\circ$, and becomes negative below that angle though it should be a positive quantity by definition. Therefore the spectrum obtained in Ref.~\cite{dosSantos} seems valid at most in the region $\xi\gg 3$. At any rate, the LL spectrum according to the Fermi speed renormalization shown in Ref.~\cite{dosSantos}
is given by
\begin{align}
E_{\ell}=\pm v_{\rm F}\left(1-9\frac{\tilde t_\perp^2}{\hbar^2 v_{\rm F}^2 |\Delta{\bf K}|^2}\right){\sqrt{2\hbar eB\ell}}~,\cr (\ell=0,1,2,\cdots)~,\label{dS}
\end{align}
that is, the LL for massless fermions modulated by the renormalized Fermi speed.

\begin{figure}[!ht]
\includegraphics[scale=0.9]{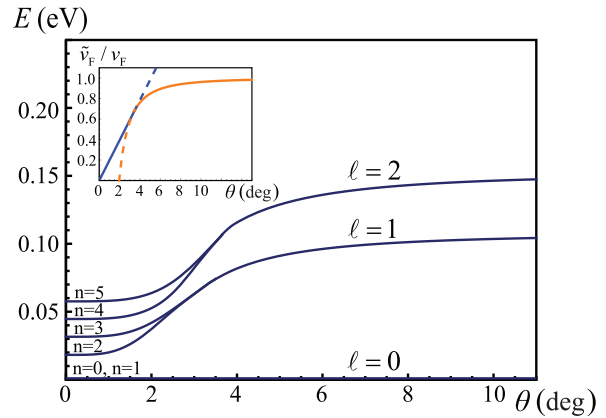}
\caption{\small (Color online) The qualitative behavior of the LL's ($B=10\,$T) as functions of $\theta$, modulated by the renormalized Fermi speed. Inset: expected behavior of the renormalized Fermi speed, interpolating \eqref{vf1} and \eqref{vf2} (the solid lines).}
\label{fig:LL}
\end{figure}

Note, however, that the validity regions for the spectra \eqref{Eei} and \eqref{dS} can have some overlap for $3\ll \xi \ll 15$. Actually, as we remarked above,
both the spectra at an intermediate angle such as $\theta=5^\circ$ are of massless character. Except the modulation function $\hbar\omega(\theta)$ depending on the renormalized Fermi speed, they are the same unless the applied magnetic field is very very large.
The renormalized Fermi speed according to \eqref{twobandH} and \eqref{largeB} (see also the caption of Fig.~\ref{fig:band}) is, to the first order in $\xi/15$, linear in $\xi$:
\begin{align}
\frac{\tilde v_{\rm F}}{v_{\rm F}}=\frac{2\hbar v_{\rm F} |\Delta{\bf K}|}{15\,\tilde t_\perp}=\frac{2}{15}\xi~.\label{vf2}
\end{align}
All these circumstances make it plausible that the spectra \eqref{Eei} and \eqref{dS} are smoothly connected in the intermediate range of $\xi$ (or $\theta$) and, here comes the punchline, the curves representing the renormalized Fermi speed for these two cases are almost tangent to each other, at the angle $\theta\simeq3.37^\circ$ ($\xi\simeq 5.20$). The linear function $f(\xi)$ which is exactly tangent to the curve $\tilde v_{\rm F}/v_{\rm F}=1-9/\xi^2$ is $f(\xi)=2\xi/9\sqrt3$. See the inset in Fig.~\ref{fig:LL}.
If the renormalizaion of the Fermi speed is given by the curve depicted here, the LL's smoothly connecting \eqref{Eei} and \eqref{dS} should
behave qualitatively as Fig.~\ref{fig:LL}. The Fermi speed renormaliztion affects the spectrum for angles $\theta\gtrsim 3.37^\circ$ to change the shape of its
tail.

Recall that the values of $\theta_{\rm c,deg}$ predicted in Sec.~\ref{sec3} are independent of the (renormalized) Fermi speed since their predictions are based only on the spheroidal eigenvalues themselves, i.e., $E_n/\hbar\omega$. Therefore, if the discussion made in this appendix based on physical continuity of the spectrum is correct, the predicted values can remain trustable. In other words, if one can find an exact formula for renormalized Fermi speed as a function of $\theta$ which interpolates the two asymptotic forms \eqref{vf1} and \eqref{vf2} (equivalently the modulation function $\hbar\omega(\theta)$), the exact LL spectrum is given by the spheroidal eigenvalues modulated by it, \eqref{Eei}, at {\it any} values of $\theta$. Thus it is extremely interesting to find such a exact interpolating function for the renormalized Fermi speed.

A final remark is in order. Since the validity of the asymptotic form \eqref{vf1} is uncertain in the intermediate region, the slope in the other asymptotic form \eqref{vf2} is also uncertain accordingly. Still the inclusion of the factor $3$ in \eqref{simplet} seems crucial, because otherwise the slope in \eqref{vf2}
must be modified into a too large number to have a chance of smooth interpolation. The other numerical factor $5/2$ in \eqref{simplet} is the actual source of uncertainty, and it can be replaced by a number in some range --- roughly from 2 to 2.5. If we adopted the simplified interlayer coupling
\begin{align}
\sum H_\perp \rightarrow -3\times 2\tilde t_\perp \begin{pmatrix} 0 & 1 \\ 0 & 0 \end{pmatrix}
\end{align}
instead of \eqref{simplet}, the slope of \eqref{vf2} becomes $1/6$ which is a reasonable number to make the interpolation. Actually, the interlayer coupling matrix
$H_\perp=-2\tilde t_\perp \begin{pmatrix} 0 & 1 \\ 0 & 0 \end{pmatrix}$ approximates each of the coupling terms in \eqref{hperp} much more closely, as one can see by performing the diagonalization of \eqref{fullH} after turning off the block diagonal Dirac Hamiltonians. The coupling constant $\tilde t_\perp$ should be close to $\gamma_1/2$ in this case.

\end{appendix}

\end{document}